\documentclass[preprint,showpacs,aps,prb]{revtex4}
\usepackage{amssymb,amsmath}
\usepackage{graphicx,float}
\usepackage{bm}
\begin{document}
\title{Electrocaloric effect in relaxor ferroelectrics}

\author{R. Pirc}
\email{rasa.pirc@ijs.si}
\author{Z. Kutnjak}
\author{R. Blinc}
\affiliation{Jo\v zef Stefan Institute,  P.O. Box 3000, 1001 Ljubljana, 
Slovenia}
\author{Q. M. Zhang}
\affiliation{Materials Research Institute and Electrical Engineering 
Department,\\
The Pennsylvania State University, University Park, PA 16802, USA}

\date{\today}

\begin{abstract}

The electrocaloric effect (ECE) in normal and relaxor ferroelectrics 
is investigated in the framework of a thermodynamic approach based on
the Maxwell relation and a Landau-type free energy model. 
The static dielectric response of relaxors is described by the 
spherical random bond--random field model, yielding the first Landau
coefficient $a=a(T)$, which differs from the usual expression
for ferroelectrics. The fourth-order coefficient $b$ is treated as
a phenomenological parameter, which is either positive or negative due
to the anisotropy of the stress-mediated coupling between the polar 
nanoregions. When $b<0$, the maximum ECE in a relaxor is predicted
near the critical point in the temperature-field phase diagram, 
whereas in a ferroelectric it occurs at the first order phase 
transition. The theoretical upper bound on the ECE 
temperature change is estimated from the values of saturated 
polarization, effective Curie constant, and specific heat of 
the material. 

\end{abstract}


\maketitle
~~~~


\section{Introduction}

Several articles have recently focused on the electrocaloric effect (ECE)
in ferroelectrics and related materials,\cite{M1,M2,C1,N1} which bears  
analogy with the well known magnetocaloric effect (MCE).\cite{TS,GPT} 
Here we investigate the mechanisms of ECE in relaxor ferroelectrics, 
to be referred to as {\it relaxors} and normal ferroelectrics (or 
{\it ferroelectrics}), and discuss the specific features of these two 
groups of materials. In particular, we will discuss the possibility
of achieving a giant ECE in bulk inorganic relaxors and ferroelectric 
materials as well as in organic polymers. These systems offer 
the prospect of practical applications, such as miniaturized and 
energy efficient cooling devices, without the need for large 
electric currents commonly associated with the MCE. 

A crucial physical quantity in ECE is the change of entropy of
a polar material under the application and/or removal of an external 
electric field. For example, when the electric field is turned on 
isothermally, the elementary dipolar entities in the system will
become partially ordered and the entropy will be lowered. The
entropy lowering of the dipolar subsystem is then compensated by 
an increase of the temperature of the total system, which characterizes 
the ECE. The degree of lowering depends on the number of 
statistically significant configurations in the initial and final states 
of the system, as well as on the size of the average dipole moment and 
the volume density of dipolar entities. Other factors may also play a 
role: If the system undergoes a first order phase transition under 
the action of external electric field, the entropy will be enhanced on 
crossing the borderline between the two phases, resulting in a larger 
ECE. The line of first order transition points terminates at a critical
point where the transition becomes continuous,\cite{K5} and it will 
be of special interest to investigate the behavior of ECE in the 
vicinity of the critical point.

Estimates of the ECE can be made on the basis of thermodynamic Maxwell 
relations using the measured heat capacity and the field and temperature 
dependence of the dielectric polarization. From the theoretical point 
of view, a central problem is how to make predictions about the 
temperature and field dependence of ECE. As a first step, one needs 
to develop an appropriate phenomenological and/or mesoscopic model, 
which incorporates the specific physical features of the systems. 
Here we will make use of the standard Landau phenomenological 
model, which can be applied to both relaxors and ferroelectrics with 
the corresponding choice of Landau coefficients. These in turn can be 
derived from the mesoscopic model of the material under study. In 
the case of relaxors, the mesoscopic model of choice is the spherical 
random bond--random field (SRBRF) model, which is based on the concept 
of reorientable polar nanoregions (PNRs).\cite{PB1} Thus we should be
able to compare the ECE in relaxors and ferroelectrics, and determine 
the parameters, which control the ECE in these systems. Finally, 
using general principles of statistical thermodynamics we will discuss
the existence of a theoretical upper bound on the ECE and argue 
that it satisfies a universal relation, which is, in principle, also 
applicable to MCE.


\section{Thermodynamic approach}

The temperature change of a polar system under adiabatic electric 
field variation from the initial value $E_i=0$ to final 
value $E_f=E$ can be written in the form \cite{M1}
\begin{equation}
\Delta T =-\frac{T}{C_E}\int_0^E 
\left(\frac{\partial P}{\partial T}\right)_{\!E} dE,
\label{DT1}
\end{equation}
which follows from the thermodynamic Maxwell relation  
$(\partial S/\partial E)_T=(\partial P/\partial T)_E$ 
involving the entropy density $S(E,T)$ and the physical dielectric 
polarization $P$ (in units of C/m$^2$). The volume specific heat 
at constant field is given by $C_E = \rho c_E$.

In deriving Eq.~(\ref{DT1}), 
one tacitly assumes that the fluctuations of polarization 
$P({\vec r})$ can be ignored and that $P$ represents a thermodynamic 
variable given by the macroscopic average of $P({\vec r})$.
Furthermore, it is implied that the system is ergodic, i.e.,
its response time much shorter than the experimental time scale.  

If the field and temperature dependence of $P(E,T)$
is known from experiments, the integral in Eq.~(\ref{DT1}) can be 
evaluated, yielding an estimate for $\Delta T$.\cite{C1,N1}

In model calculations, it seems convenient to change the 
integration variable in Eq.~(\ref{DT1}) from $dE$ to $dP(E)$.
This is readily done by applying the thermodynamic identity\cite{EHS}
\begin{equation}
\left(\frac{\partial P}{\partial T}\right)_{\!E}
=-\left(\frac{\partial E}{\partial T}\right)_{\!P}
\left(\frac{\partial P}{\partial E}\right)_{\!T},
\label{PTE}
\end{equation}
with the result
\begin{equation}
\Delta T = \frac{T}{C_E}\int_{P_0}^P 
\left(\frac{\partial E}{\partial T}\right)_{\!P} dP.
\label{DT2}
\end{equation}
This expression is fully equivalent to Eq.~(\ref{DT1}), with
the new integration limits given by $P_0=P(0,T)$ and $P=P(E,T)$. 

The partial derivative $(\partial E/\partial T)_P$ 
can be obtained from the free energy density functional $F(P,T)$. 
Ignoring fluctuations of the order parameter $P$, 
we write $F$ as a power series
\begin{equation}
F=F_0+\frac{1}{2}\,a\,P^2
+\frac{1}{4}\,b\,P^4+\frac{1}{6}\,c\,P^6+\cdots -EP.
\label{F}
\end{equation}
This has the standard form of a mean field free energy expansion
with temperature dependent coefficients $a, b, c,...$, etc. 

Applying the equilibrium condition $(\partial F/\partial P)_T=0$, 
we obtain the equation of state
\begin{equation}
E=a\,P+b\,P^3+c\,P^5+\cdots , 
\label{es1}
\end{equation}
and the temperature derivative in Eq.~(\ref{DT2}) becomes
\begin{equation}
\left(\frac{\partial E}{\partial T}\right)_{\!P}=
a_1\,P+b_1\,P^3+c_1\,P^5+\cdots , 
\label{ET}
\end{equation}
where $a_1\equiv da/dT$, $b_1\equiv db/dT$ etc. 
It should noted be that $P=P(E,T)$ in Eq.~(\ref{DT2}) is that solution of 
Eq.~(\ref{es1}), which simultaneously minimizes the free energy (\ref{F}).

The integration in Eq.~(\ref{DT2}) can now be carried out, yielding
\begin{equation}
\Delta T = \frac{T}{C_E}
\left[\frac{1}{2}\,a_1\,\left(P^2-P_0^2\right)
+\frac{1}{4}\,b_1\,\left(P^4-P_0^4\right)
+\frac{1}{6}\,c_1\,\left(P^6-P_0^6\right)+\cdots \right].
\label{DT3}
\end{equation}

In passing, we note that $C_E$, in general, depends on the temperature; 
however, in writing down Eqs.\ (\ref{DT1}) and (\ref{DT2}) the temperature 
dependence of the heat capacity had already been ignored.

The expression in brackets is related to the change of the entropy 
density $S=-(\partial F/\partial T)_E$. Using Eq.~(\ref{F}) we can write
\begin{equation}
S=S_0+S_1(P).
\label{S}
\end{equation}
The first term $S_0=-\partial F_0/\partial T$ is the entropy at $P=0$.
It contains the configuration entropy of dipolar 
entities, which depends on the number of equilibrium orientations 
$\Omega$, say, $\Omega=8$ for the $<111>$ equilibrium case.\cite{M3}  Thus 
we may expect that $S_0\sim (N/V)k\ln (\Omega)$, $N$ being the total 
number of dipolar entities such as PNRs in relaxors. The second term 
is given by 
\begin{equation}
S_1(P)=-\left(\frac{1}{2}\,a_1\,P^2+\frac{1}{4}\,b_1\,P^4
+\frac{1}{6}\,c_1\,P^6+\cdots\right) .
\label{S1}
\end{equation}
Back in Eq.~(\ref{DT3}), $S_0$ cancels out and the ECE temperature 
change can be rewritten in the familiar form \cite{TS}
\begin{equation}
\Delta T=-\frac{T}{C_E}\Delta S_P,
\label{DT4}
\end{equation}
with $\Delta S_P\equiv S_1(P)-S_1(P_0)$.

It should be noted that the values of all temperature-dependent 
quantities $P_0$, $P$, $a_1$, $b_1$, etc., on the r.h.s. 
of Eq.~(\ref{DT3}) are taken at the initial temperature $T_i=T$, 
and $C_E$ at the final field value $E$. 

The coefficients $a,b,c,...$ can be expressed in terms of linear
and nonlinear susceptibilities by formally inverting the relation 
(\ref{es1}) and writing $P(E)$ as a power series in $E$.\cite{I2}
In Landau theory, close to a second order phase transition one sets 
$a(T)\propto T-T_0$, while $b,c,...$ are constants. Thus, 
$a_1=$~const., and $b_1=c_1=\cdots=0$. This leaves only one nonzero 
term of the order ${\cal O}(P^2)$ in Eq.~(\ref{DT3}). On the other 
hand, $\chi_1\propto |T-T_0|^{-1}$ and the nonlinear susceptibilities 
are also found to diverge when $T\to T_0$. Thus a formal inversion 
$P(E)$ of Eq.~(\ref{es1}) in powers of $E$ would lead to a poorly 
converging series.
 
In the following we will apply Eq.~(\ref{DT3}) in order to discuss 
the predictions of Landau theory in two characteristic cases, namely, 
normal ferroelectrics and relaxors.


\section{SRBRF model of ECE in relaxors and ferroelectrics}

As already mentioned, in Landau theory of phase transitions in
ferroelectrics, the coefficients $b,c,...$ in Eq.~(\ref{F}) 
are assumed temperature independent and $a\propto T-T_0$, where $T_0$ 
is the Curie-Weiss temperature. When $b>0$ and $E=0$, a second order 
transition occurs at $T_c=T_0$. For $b<0$ and $c>0$, a first order 
transition appears at a temperature $T_1$, given by the relation 
$a(T_1)=3b^2/16c$. 
Writing $a=(T-T_0)/(\varepsilon_0\Theta)$, where $\Theta$ is the Curie 
constant, we find:
\begin{equation}
T_1=T_0+ \varepsilon_0\Theta\frac{3}{16}\frac{b^2}{c}.
\label{t1} 
\end{equation}
For $E\ne 0$, a critical point will be located at $T_{CP},E_{CP}$,
where \cite{I1}
\begin{equation}
T_{CP}=T_0+ \varepsilon_0\Theta\frac{9}{20}\frac{b^2}{c}, \;\;\;\;\;\;
E_{CP}=\frac{6b^2}{25c}\sqrt{\frac{3|b|}{10c}}.
\label{CP} 
\end{equation}

Turning next to relaxors, we assume that the relevant elementary 
dipolar entities at temperatures around the dielectric maximum 
are polar nanoregions or PNRs. 
According to the SRBRF model,\cite{PB1} these PNRs are coupled through 
Gaussian random interactions $J_{ij}$  ("random bonds") and are subject 
to Gaussian random fields $h_i$.
In zero applied field, spontaneous long range order is suppressed 
$(P_0=0)$. This means that for a relaxor we can still use the free 
energy (\ref{F}), however, the coefficient $a(T)$ must remain positive 
at all temperatures. Thus, for $b>0$ and $E=0$ there can be no second 
order phase transition. The explicit form of $a(T)$ follows from 
Eq.~(\ref{es1}), namely, 
\begin{equation}
a=\left(\frac{\partial E}{\partial P}\right)_{P=0}=
(\varepsilon_0\chi_1)^{-1},
\label{achi} 
\end{equation}
where $\chi_1=(\partial P/\partial E)_{E=0}$ represents 
the (quasi)static linear field-cooled dielectric susceptibility. 
This can be derived from the SRBRF model of relaxors:\cite{PB1}
\begin{equation}
\chi_1=\frac{\Theta(1-q)}{T-T_0(1-q)},
\label{chi1} 
\end{equation}
with 
\begin{equation}
\Theta=\frac{g^2}{vk\varepsilon_0}.
\label{th} 
\end{equation}
The effective Curie constant $\Theta$ is given explicitly in terms 
of the average squared dipole moments $g_i$ of PNRs, namely, 
$g^2=\sum_{i=1}^Ng_i^2/N$, and the average
volume $v=V/N$ associated with a PNR. Later, we will also introduce
the saturation polarization $P_{max}={\bar g}/v$, where 
${\bar g}\equiv \sum_{i=1}^Ng_i/N$. For simplicity, we will 
henceforth neglect the difference between ${\bar g}$ and $g$, 
and write $P_{max}\cong g/v$.

The parameter $\Theta$ can be determined experimentally 
from the asymptotic high temperature behavior of $\chi_1$.\cite{U1,V1}
The parameter $T_0=J_0/k$ is defined {\it via} the average over  
the infinitely ranged random interaction $[J_{ij}]_{av}=J_0/N$. The 
spherical glass order parameter, $0<q<1$, is a measure of 
the degree of disorder. For $P=0$ it is determined by the cubic equation
\begin{equation}
(kT)^2q=(J^2q+\Delta)(q-1)^2.
\label{q} 
\end{equation}
Here, $J^2$ is proportional to the variance of the random bond 
distribution according to $[J_{ij}^2]_{av}-([J_{ij}]_{av})^2=J^2/N$, 
while $\Delta$ measures the correlations of quenched random fields,   
i.e., $[h_i]_{av}=0$ and $[h_ih_j]_{av}=\delta_{ij}\Delta$. 
If $J_0<\sqrt{J^2+\Delta}$, long range order will be suppressed ($P_0=0$). 
Thus the relaxor state is characterized by three physical parameters: 
$J_0$, $J$, and $\Delta$. Typically, in relaxors one finds $\Delta \ll J^2$, 
implying that random bonds are effectively much stronger than random fields. 
This allows PNRs to reorient collectively under the action of external 
fields and relax towards equilibrium. In the opposite case, 
$\Delta \gg J^2$, PNRs are would be trapped in a frozen static 
configuration of random fields and no characteristic low-frequency 
relaxor response due to PNR flipping could be observed. 
Here we will consider only the case $\Delta \ll J^2$.

From Eqs.\ (\ref{chi1}) and (\ref{achi}) we derive the coefficient $a(T)$:
\begin{equation}
a(T)=\frac{1}{\varepsilon_0\Theta}\left(\frac{T}{1-q}-T_0\right).
\label{at1} 
\end{equation}

The linear susceptibility $\chi_1$ has been fitted to experimental data 
for the static field-cooled response in a variety of relaxor systems,
from which the parameters of the model have been 
obtained.\cite{K1,P2,B1} 

Formally, the SRBRF model 
also yields explicit expressions for nonlinear susceptibilities, from 
which the coefficients $b$ and $c$ in the free energy (\ref{F}) can be 
determined.\cite{I2}  However, it has been shown earlier\cite{K1,P2,PBV} 
that realistic values of these coefficients can only be 
obtained if the coupling between PNRs and lattice strain fluctuations 
is included. Several mechanisms for such a coupling have 
been investigated both at a mesoscopic \cite{K1,P2,PBV} and 
phenomenological level.\cite{PBS} It has also been shown\cite{PBV,PBS} 
that in real three dimensional systems strain coupling gives rise 
to anisotropy of the anharmonic terms in the free energy.\cite{PBV} 
Specifically, strain coupling may change the sign of the 
coefficient $b$ and hence of the corresponding nonlinear 
susceptibility $\chi_3$ for a given direction of the applied field. 
In the following, we will simply consider $b$ and $c$ as free parameters 
and discuss separately the cases $b>0$ and $b<0$, while keeping $c>0$.

For cubic systems, Eq.~(\ref{F}) can be rewritten in a
general form \cite{ANC}
\begin{equation}
\begin{split}
&F=\alpha_1\sum_i P_i^2+\alpha_{11}\sum_i P_i^4
+\alpha_{12}\sum_{i\ne j} P_i^2P_j^2+\alpha_{111}\sum_i P_i^6\\ 
&+\alpha_{122}\sum_{i\ne j\ne k} 
P_i^4(P_j^2+P_k^2)
+\alpha_{123}P_1^2P_2^2P_3^2-\sum_iE_iP_i,
\end{split}
\label{Fc1}
\end{equation} 
where $\alpha_i,\alpha_{ij},\alpha_{ijk}$ are so-called dielectric 
stiffness coefficients with $i,j,k=1,2,3$.
For a system with orthorhombic symmetry in a field along  
$[111]$ we recover Eq.~(\ref{F}), where $P$ is the total polarization, 
$a=2\alpha_1$, $b_{[111]}=(4/3)(\alpha_{11}+\alpha_{12})$, and 
$c_{[111]}=(2/9)(3\alpha_{111}+6\alpha_{112}+\alpha_{123})$.\cite{QJ}
Experiments in PMN \cite{K1,K2} indicate that $\chi_3>0$ for 
${\vec E}\parallel [111]$, implying $b_{[111]}<0$. On the other hand, 
for tetragonal symmetry with ${\vec E}\parallel [001]$ one has 
$b_{[001]}=4\alpha_{11}$ and $c=6\alpha_{111}$. For example,
in PMN \cite{K1,K2} one finds in this case that $\chi_3<0$ or 
$b_{[001]}>0$, indicating that $\alpha_{11}>0$ and 
$\alpha_{12}<-\alpha_{11}$.

If $a(T)>b^2/c$ at all temperatures, there is no first order phase 
transition for $E=0$. If $E>0$, however, a first order phase transition 
with a jump of polarization $\Delta P$ occurs for $E$ exceeding some 
threshold value $E_1$. This can readily be seen by numerical minimization 
of the free energy (\ref{F}) for any pair $E,T$. As the temperature 
increases, $\Delta P(E,T)$ decreases and vanishes at an isolated 
critical point $T_{CP},E_{CP}$,\cite{K5,K2} where the derivative 
$(\partial P/\partial T)_E$ diverges.\cite{I1} 
The critical temperature is determined from the equation
\begin{equation}
a_{CP}\equiv a(T_{CP})=\frac{9}{20}\frac{b^2}{c},
\label{tcp} 
\end{equation}
and $E_{CP}$ from the second of Eqs.~(\ref{CP}). It should be 
noted that, generally speaking, $b$ and $c$ can also be 
functions of $T$.\cite{K5,K2}

For $E>E_{CP}$ the system is in a supercritical regime with 
continuous temperature and field dependence of $P(E,T)$.


\section{Temperature and electric field dependence of ECE}

To illustrate the temperature and field dependence of the ECE in
relaxors and ferroelectrics, we calculate $\Delta T(E,T)$ from 
Eq.~(\ref{DT3}) for a selected set of parameter values. First, 
we introduce rescaled, dimensionless quantities $F$ and $P$ according 
to $F\to Fv/J$ and $P\to P/P_{max}$, where $P_{max}\cong g/v$ is the 
saturation polarization occurring at high field values and/or 
low temperatures. This requires a rescaling of the remaining parameters
according to $a\to av/(JP_{max}^2)$, $b\to bv/(JP_{max}^4)$, 
$c\to cv/(JP_{max}^6)$, etc. Also, we redefine 
$T\to kT/J$ and $E\to gE/J$. From Eq.~(\ref{at1}) 
we see that in relaxors the rescaled parameter $a(T)$ behaves as 
$a(T)=T/(1-q)-T_0$, and $a_1(T)$ becomes
\begin{equation}
a_1(T)=\frac{\partial}{\partial T}\left(\frac{T}{1-q}\right).
\label{a1} 
\end{equation}

Here and until the end of this section, the symbols $a,b,c,...$ and 
$P,T,E$ refer to dimensionless, rescaled parameters, but elsewhere 
in this paper the same symbols denote the true, physical values of 
these quantities.

In ferroelectrics, $q=0$ and thus $a(T)=T-T_0$, and $a_1=1$. In relaxors,
in the high temperature limit $T>>1$, $q$ tends asymptotically to zero, 
thus $a\sim T-T_0$ and $a_1\sim 1$, i.e., the same as in ferroelectrics. 

In Fig.~1, $a(T)$ and $a_1(T)$ are plotted
for a relaxor ferroelectric with $J_0/J=0.9$ and $\Delta/J^2=0.001$. 
Also shown is the behavior of a normal ferroelectric. It should be 
noted that the essential difference between ferroelectrics 
and relaxors is the behavior of the corresponding coefficients 
$a(T)$ and $a_1(T)$. 

In the following we will choose $b=$~const.~$=\pm1/3$ and $c=|b|$. The 
corresponding ECE temperature change is obtained from Eq.~(\ref{DT3}). 
Using the fact that $b_1=c_1=0$ etc., we find
\begin{equation}
\Delta T=\frac{kT}{2vC_E}a_1(T)\Bigl[P(E,T)^2-P_0(T)^2\Bigr].
\label{DT5} 
\end{equation}
The polarization $P(E,T)$ will be  
calculated numerically by simultaneously solving Eq.~(\ref{es1}) and 
minimizing the free energy (\ref{F}). We will do that 
separately for the two cases $b>0$ and $b<0$, assuming
the denominator $2vC_E$ to be a constant amplitude factor.

(i) {\it Case $b>0$}. As already stated, $P_0=P(0,T)=0$ in a 
relaxor, but in ferroelectrics $P_0(T)\ne 0$ for $T<T_0$. The 
spontaneous polarization $P_0(T)$ is obtained by minimization of 
$F(E=0)$. In real systems, $P_0$ may not be spatially uniform due to
domains. Here we assume that $|{\vec P}_0|$ has the same value in all 
domains regardless of their orientation, and that the contribution 
of domain walls to the entropy can be neglected.\cite{TS} 

In Fig.~2(a) we show the calculated values of $\Delta T$ for a relaxor 
as function of temperature for various values of $E/E_{CP}$, where
$E_{CP}$ is formally given by Eq.~(\ref{CP}), although
the critical point does not exist for $b>0$. Also shown in Fig.~2(b) 
is $\Delta T$ for a ferroelectric with the same parameters $b, c$, 
but with different $a$ and $a_1$. $\Delta T$ has a peak at $T_0=0.9$ and 
is in general larger than in the relaxor case. At higher temperatures,
however, the difference gradually disappears.

(ii) {\it Case $b<0$}. Eqs.\ (\ref{chi1}-\ref{q}) imply 
$a(0)=\sqrt{1+\Delta/J^2}-J_0/J\cong 0.1005$. Thus, $a(0)<3b^2/16c$
and the first order phase transition in relaxors at $E=0$ is 
suppressed. In a ferroelectric, however, a first order transition 
in zero field occurs at $T_1=0.9625$. 
The critical point is located at $T_{CP}=1.0398$ in relaxors,
and at $T_{CP}=1.015$ in ferroelectrics, while $E_{CP}=0.0438$
in both cases. In general, $\Delta T$ is found to increase with 
increasing field and exhibits a peak as a function of temperature. 
In relaxors, the peak position moves to higher temperatures with 
increasing field values, whereas in ferroelectrics the maximum is 
located at $T=T_1$, where a jump of the spontaneous polarization occurs.  

In Fig.~3, $\Delta T(T)$ is plotted for four values of the field
$(E/E_{CP})^2$, as indicated. At $E=E_{CP}$ in relaxors, there is a
jump of $\Delta T$ due to the field-induced first-order transition.
On the other hand, in the case of a ferroelectric, there is a 
jump of $\Delta T$ at the zero-field first order transition temperature 
$T_1=0.9625$. At the critical point, $\Delta T$ is continuous
but with an infinite slope in both cases, while at higher temperatures 
the difference between relaxors and ferroelectrics tends to disappear. 

In Fig.~4, the ECE efficiency $\Delta T/E$ is plotted as a function of 
$E/E_{CP}$ for four values of temperature close to $T_{CP}$. As
expected, the maximum efficiency is obtained at the corresponding 
critical points $T=T_{CP}$ and $E=E_{CP}$.

Larger values of $\Delta T$ in ferroelectrics rather than in relaxors 
are mainly due to the sharp decrease of $a_1(T)$ in relaxors
at $T\lesssim 1$ (cf. Fig.~1). However, this does not mean that 
ferroelectrics are better candidates for achieving giant ECE. 
Namely, one should bear in mind that there are other parameters, 
such as $\Theta$ and $C_E$, which also have a strong impact on the ECE. 
Moreover, the above comparison between relaxors and ferroelectrics 
makes only sense if the coefficients $b, c,...$ are indeed the same 
in both cases. Therefore, in discussing the ECE in specific systems 
one should carefully consider the actual physical values of all the 
relevant model parameters, as discussed in the following section. 
 

\section{Application to real systems}

It has been observed experimentally in a variety of systems that 
the entropy change $\Delta S_P$ in Eq.~(\ref{DT4}) is 
proportional to $P^2$,\cite{LG,LZ,N2} suggesting that the terms 
of order $\sim P^4$ and higher in the expansion (\ref{S1}) make
no contribution. Assuming for simplicity that $P_0=0$, we recover 
from Eq.~(\ref{S1}) the empirical quadratic relation 
\begin{equation}
\Delta S_P=-\frac{1}{2}\beta P^2,
\label{DS} 
\end{equation}
where the coefficient $\beta$ can be expressed through Eqs.\ (\ref{S1}) 
and (\ref{at1}), i.e.,
\begin{equation}
\beta=a_1(T)=\frac{1}{\varepsilon_0\Theta}\frac{\partial}{\partial T}
\left(\frac{T}{1-q}\right).
\label{beta} 
\end{equation}
According to the Landau model, in ferroelectrics the partial 
derivative is equal to $1$ or $\beta = (\varepsilon_0\Theta)^{-1}$  at 
all temperatures, whereas in relaxors it approaches the value $\sim 1$ 
at high temperatures, but is in general a function of temperature. 
Thus, in relaxors, $\beta$ is expected to be a function of temperature 
with $\beta(T)\leq (\varepsilon_0\Theta)^{-1}$. 

Using the relations (\ref{DT4}) and (\ref{DS}), the ECE temperature 
change $\Delta T$ can be written as
\begin{equation}
\Delta T=\frac{T}{2C_E}\beta P(E,T)^2.
\label{DT6} 
\end{equation}

A quadratic relation of the same form is predicted by Eq.~(\ref{DT5})
if the terms $b_1$, $c_1$,... etc. are neglected. It is exactly
true in the case $b_1=c_1=\cdots=0$ discussed in Section III.

Clearly, the above relation is applicable only in the {\it quadratic}
regime where the empirical Eq.~(\ref{DS}) is valid. 
The parameter $\beta$ can be determined directly from the measured ECE 
temperature change $\Delta T$, with $P(E,T)$ extracted from dielectric
experiments. Alternatively, and especially in ferroelectrics or relaxors 
at temperatures above the freezing temperature, we can obtain both 
$P(E,T)$ and the Curie constant $\Theta$ from the dielectric data, and 
then deduce $\beta$ from Eq.~(\ref{beta}). We can then predict 
$\Delta T$ from Eq.~(\ref{DT6}). It turns out, however, that the 
values of $\Delta T$ measured directly may differ from the ones 
deduced from dielectric data. For example, in the case of relaxor 
ferroelectric terpolymer P(VDF-TrFE-CFE) at $T\sim 300$~K and 
$E=70$~MV/M, the ECE measured directly is $\Delta T \sim 3.6$~K, 
leading to $\beta\sim ~10^7$~V m C$^{-1}$K$^{-1}$,\cite{N2}
whereas the value deduced from dielectric data is  
$\Delta T \sim 0.87$~K.\cite{L1} A tentative explanation 
of this discrepancy is that
even far above the freezing temperature $T_f\sim 277$~K the system
may still be nonergodic, so that the Maxwell relations, based on
equilibrium thermodynamics, are not applicable.\cite{L1} As a
consequence, the value of $P(E,T)$ measured on a short time scale 
is smaller than its thermodynamic long-time limit. Another 
possibility is that the empirical relation Eq.~(\ref{DS}) is an 
effective quasi-linear relation observed in a broad range of large 
field values, while its derivation based on the Landau expansion is 
by assumption restricted to small fields.

We can obtain an estimate for the maximum ECE temperature change 
$(\Delta T)_{max}$ by assuming that in a sufficiently strong electric 
field the polarization reaches its saturation value $P_{max}\cong g/v$. 
Relation (\ref{DS}) is not expected to be valid in this {\it saturation}
regime, and we must return to the general expression (\ref{DT4}).
Obviously, the expansion (\ref{S1}) cannot be applied due to convergence
problems. On the other hand, it is well known \cite{M3} that in 
the saturation regime the excess entropy of the dipolar subsystem 
tends to zero. Therefore, according to Eq.~(\ref{S}), $S_1(P_{max})$ 
should approach the negative value of the configuration entropy 
$S_0$, i.e., $S_1(P_{max})\to -(k/v)\ln(\Omega)$. Thus, in the 
saturation regime Eq.~(\ref{DT4}) leads to
\begin{equation}
(\Delta T)_{max}=\frac{kT\ln(\Omega)}{v C_E}.
\label{DTm1} 
\end{equation}
This relation gives the theoretical upper bound on ECE in terms of 
just three physical quantities, $v$, $C_E$, and the configuration 
number $\Omega$. Interestingly, this result does not depend 
explicitly on the dipole moment $g$. Moreover, it does not contain 
any information about possible phase transition occurring in 
the quadratic regime. 

For a given value of electric field $E$, the borderline between
the two regimes is expected to occur at some temperature $T^*(E)$ 
where the dipolar energy becomes equal to the thermal 
fluctuation energy, i.e., $T^*\cong gE/k$. For $T\gtrsim T^*$ the
system is in the quadratic, and for $T\lesssim T^*$ in the saturation
regime. For the above terpolymer P(VDF-TrFE-CFE), we can estimate 
the dipole moment $g$ from the relations $P_{max}\sim 0.1$~C/m$^2$ 
and $g=k/(\beta P_{max})$. Using the value of $\beta$ determined 
directly from $\Delta T$, we have $g\sim 5.7\times 10^{-30}$~C m,
and for $E=300$~MV/m we thus find $T^*\sim 124$~K. 

The average volume associated with a PNR in relaxors, $v=V/N$, 
in Eq.~(\ref{DTm1}) is not {\it a priory} known and depends 
on the total number of PNRs. We can estimate $v$ from
the measured values of $\Theta$ and $P_{max}$ using the relation
$\Theta \cong P_{max}^2 v/k\varepsilon_0$, and rewrite Eq.~(\ref{DTm1})
in the form
\begin{equation}
(\Delta T)_{max} \cong \frac{T\ln(\Omega)}{\varepsilon_0\Theta C_E}P_{max}^2.
\label{DTm2} 
\end{equation}
The value of $P_{max}$ can be extracted, for example, from hysteresis 
loops in the saturation regime, and $\Theta$ from the asymptotic behavior 
of $\chi_1$. As already discussed above, this value of $\Theta$ may
differ from the one derived from the experimental value of parameter 
$\beta\cong (\varepsilon_0\Theta)^{-1}$, observed in ECE experiments
in the effective quadratic regime.

In Table~I, the predicted values of $(\Delta T)_{max}$ for a set of
selected systems are listed using the values of $\Theta$ deduced from
dielectric experiments. It should be noted that Eqs.\ (\ref{DTm1}) 
and (\ref{DTm2}) provide just a theoretical upper bound for
the ECE in the systems listed.
In practice, the limit of a fully polarized dipolar subsystem might  
not be accessible because dielectric breakdown could occur before 
complete saturation is reached. Nonetheless, the predicted values 
of $(\Delta T)_{max}$ permit a comparison between various sytems 
and might be useful in the search for a giant ECE.

Eq.~(\ref{DTm1}) indicates that a giant ECE is expected to occur 
in systems with a small value of $v$, or equivalently a large number 
$N$ of dipolar entities at fixed volume $V$. For illustration, let
us consider a specific example, i.e., the ferroelectric
copolymer P(VDF-TrFE). This system consists of microscopic crystalline 
layers of polarized material embedded in an amorphous environment.\cite{QZ} 
Electron irradiation  breaks up the layered structure into smaller 
dipolar units, and turns the polymer into a relaxor. Since the number 
of these new entities is now larger than the number of the original 
microcrystallites, and assuming the same value of saturation
polarization, one expects a stronger ECE to occur in the irradiated 
relaxor copolymer than in the original ferroelectric copolymer. 
This is corroborated by the experimental values of the coefficient 
$\beta$ for the irradiated relaxor copolymer P(VDF-TrFE),\cite{L1}
$\beta\cong 9.3\times 10^7$~V m C$^{-1}$K$^{-1}$,
and for the original ferroelectric copolymer in the paraelectric 
phase,\cite{N2} $\beta\cong 5.8\times 10^7$~V m C$^{-1}$K$^{-1}$.

In the case of inorganic relaxor 8/65/35 PLZT thin films, the value 
of the coefficient $\beta$ is\cite{L1} 
$\beta \sim 1.5\times 10^6$~V m C$^{-1}$K$^{-1}$, whereas for
the ferroelectric PZT one finds\cite{M1} 
$\beta\sim 7.6 \times 10^5$~V m C$^{-1}$K$^{-1}$. Again,
$\beta$ is larger for the relaxor; however, this comparison
seems less conclusive since the difference in composition between 
the two systems is much greater than in the above organic case.  

Relation (\ref{DTm1}) has been derived here in the framework of 
Landau theory, however, its validity is essentially based on 
thermodynamic and statistical principles and is hence quite general. 
In particular, it is independent of any mesoscopic models such as 
the SRBRF model. Moreover, it can be easily generalized to magnetic 
systems, where $\Delta T$ represents the MCE temperature change.

The smallest physical limit for $v$ in ferroelectrics is the volume 
of the unit cell $v_0$, yielding the ultimate upper bound on 
$(\Delta T)_{max}$ in Eq.~(\ref{DTm1}) for ferroelectric materials.
To emphasize this point, let us multiply the numerator and 
denominator by the Avogadro number $N_A$ to obtain 
\begin{equation}
\left(\frac{\Delta T}{T}\right)_{max}=\frac{R\,\ln(\Omega)}{C_m}.
\label{DTm3} 
\end{equation}
Here, $R=N_Ak$ is the gas constant and $C_m=N_Av_0C_E$ the molar 
specific heat. This simple result does not explicitly contain any 
information on the microscopic nature of the system, i.e., whether 
it is dielectric, magnetic, etc., since the corresponding electric 
or magnetic dipole moment does not appear in Eq.~(\ref{DTm3}). 
Therefore, we can regard the above relation as a universal law 
for the theoretical upper bound on the ECE and/or MCE in 
electrically or magnetically polarizable solids. Of course, 
Eq.~(\ref{DTm3}) implies that the polarization or magnetization must
have reached complete saturation. In the dielectric case, this requires 
large electric fields with the already mentioned possibility of 
dielectric breakdown. In the magnetic case, however, extremely
large fields of the order of $\sim 100$~T may be necessary.

Eq.~(\ref{DTm3}) is valid in the saturation regime $T<T^*$,
typically at low temperatures, where $C_m$ is expected to 
be temperature dependent. In ferroelectrics, the
formation of domains should be taken into account, and the entropy 
limit $S \to -S_0$ could only be reached in very high fields. 
In relaxors, there is an additional difficulty that
the relaxation of PNRs at low temperatures is extremely slow 
and the above limit can only be reached after very long times.

At high temperatures, $C_m$ normally approaches a certain limit.
In metals one has $C_m\sim 3R$ according to the
Dulong-Petit law, which is generally not valid for complex solids.
For example, in perovskites, $C_m\sim 15 R$.\cite{M3} Thus, the 
maximum ECE in perovskites at high temperatures is of the order 
$(\Delta T/T)_{max}\sim 0.0667 \ln(\Omega)$. This limit applies to 
a large group of perovskites, i.e., paraelectrics, ferroelectrics, 
relaxors, etc. Specifically, for $\Omega=8$ (see Table I), we find
$(\Delta T/T)_{max}\sim 0.139$. For example, if the system can 
support large electric fields such that $T^*>300$~K, this would 
lead to a giant ECE temperature change $(\Delta T)_{max} \sim 40$~K 
at room temperature. This exceeds roughly by a factor of $\sim 2$ 
the estimated values for perovskites in Table~I, 
which were derived from the physical values of the parameters 
$P_{max}$, $\Theta$, and $C_E$.


\section{Conclusions}

In this paper, we studied the mechanism of ECE in relaxor ferroelectrics
(or {\it relaxors}) and in normal ferroelectrics (referred to as
{\it ferroelectrics}). Starting from the widely accepted result for
the ECE temperature change $\Delta T(E,T)$, based on the thermodynamic 
Maxwell relation, we derived an alternative expression which could
directly be applied in theoretical model calculations. The results 
for $\Delta T(E,T)$ have been obtained in two physical regimes of 
field-temperature variables $E,T$. The first of these is the so-called 
{\it quadratic} regime, where $\Delta T$ is proportional to the square 
of the dielectric polarization $P(E,T)$, and the second the 
{\it saturation} regime, where $P(E,T)$ is allowed to reach its 
maximum value $P_{max}$. 

In the quadratic regime, the system can be described by the Landau-type 
free energy model, in which the harmonic Landau coefficient $a(T)$ 
depends on the physical nature of the system, i.e., the behavior
of $a(T)$ in relaxors differs from that of ferroelectrics. The anharmonic 
coefficients $b,c,...$, which are common to both cases, then determine 
the critical behavior of the system in a field $E$. The case $b>0$
does not show any pronounced anomalies. For $b<0$,
the polarization $P$ and hence $\Delta T$ rises steeply in a relaxor
near the isolated critical point $E_{CP},T_{CP}$, whereas in a
ferroelectric, the largest effect occurs near the first order phase
transition in zero field, which is absent in the relaxor case. The ECE 
efficiency $\Delta T/E$ shows a similar behavior in both cases, and 
at higher fields the same asymptotic values are found. Of course, 
these conclusions only apply provided that all the remaining physical
features, such as the number of equilibrium orientations $\Omega$
of the elementary dipolar entities, etc., are the same in both cases.

Experimentally, the ECE entropy change in the {\it quadratic} regime 
is found to behave as $\Delta S\sim -(1/2)\beta P(E,T)^2$, which is 
trivially reproduced by the Landau model. The coefficient $\beta$ 
tends to have larger values in relaxors, leading to a stronger ECE. 
In irradiated organic polymer relaxors, this can 
be explained by the larger number of polar nanoregions (PNRs) and 
thus a smaller average PNR volume $v$. In the {\it saturation} regime, 
the entropy of the dipolar subsystem generally approaches the 
negative value of the configuration entropy, $S\sim -(k/v)\ln(\Omega)$, 
and the maximum ECE value $(\Delta T)_{max}$ thus crucially depends 
on the orientational degeneracy $\Omega$. 

\section{Acknowledgments}

This work was supported by the Slovenian Research Agency through
Grants P1-0044, P1-0125, P2-0105, J1-0155, and J1-2015. Research 
at the Pennsylvania State University was supported by the US DoE,
Office of Basic Energy Sciences, Division of Materials Science
and Engineering under Award No. DE-FG02-07ER46410.

\newpage

\newpage

\begin{table*}
\caption{\label{tab:table1} 
~Predicted limiting values of ($\Delta T$)$_{max}$ for various 
perovskite and polymer relaxor systems according to Eq. (\ref{DTm2}).}
~~~~~\\
\begin{ruledtabular}
\begin{tabular}{cccccc}
Material & $~~\Omega~~~$ & $P_{max}~~$ & $C_E$ at 300 K &
$\beta=(\varepsilon_0\Theta)^{-1}$ & $\Delta T$ (K)\\
~&~& (C/m$^2)~~$ & ($10^6$~J/m$^3$K) & ($10^5$~Vm/C K) & at 300 K\\
\hline
8/65/35 PLZT & 8 & 0.65\cite{K3} & 3.0\cite{S1} 
& 2.43\cite{K4}  & 21.4 \\
PMN-29.5PT & 8 & 0.55\cite{K5} & 2.8\cite{K5} & 3.76\cite{K5}  & 24.8 \\
PMN& 8 & 0.55\cite{K2} & 2.6\cite{M3} & 3.82\cite{V1} &  20 \\
P(VDF-TrFE) 68/32 & 8 & 0.06\cite{L2} & 2.34\cite{L2} & 
4.67\cite{L2} & 39 \\
(irradiated film) & ~ & ~ & ~ & ~ & ~ \\
\end{tabular}
\end{ruledtabular}
\end{table*}

~~~~~~~~~~~~
~~~~~~~~~~~~

\vfill\eject

\newpage

\begin{figure}
\resizebox{20.4pc}{!}{\includegraphics{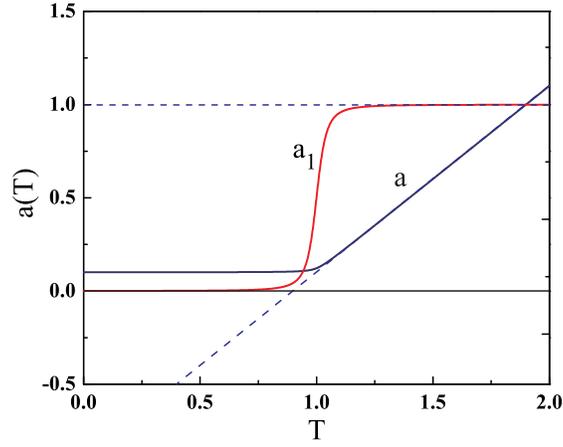}}
\caption{\label{ecc10aa}Temperature dependence of the coefficient
$a(T)$ and its temperature derivative $a_1(T)$ for a relaxor 
(solid lines) and a ferroelectric (dashed).}

\end{figure}


\begin{widetext}
\begin{figure*}
\begin{tabular}{cc} 
{\large (a)}&{\large (b)}\\
\resizebox{20.4pc}{!}{\includegraphics{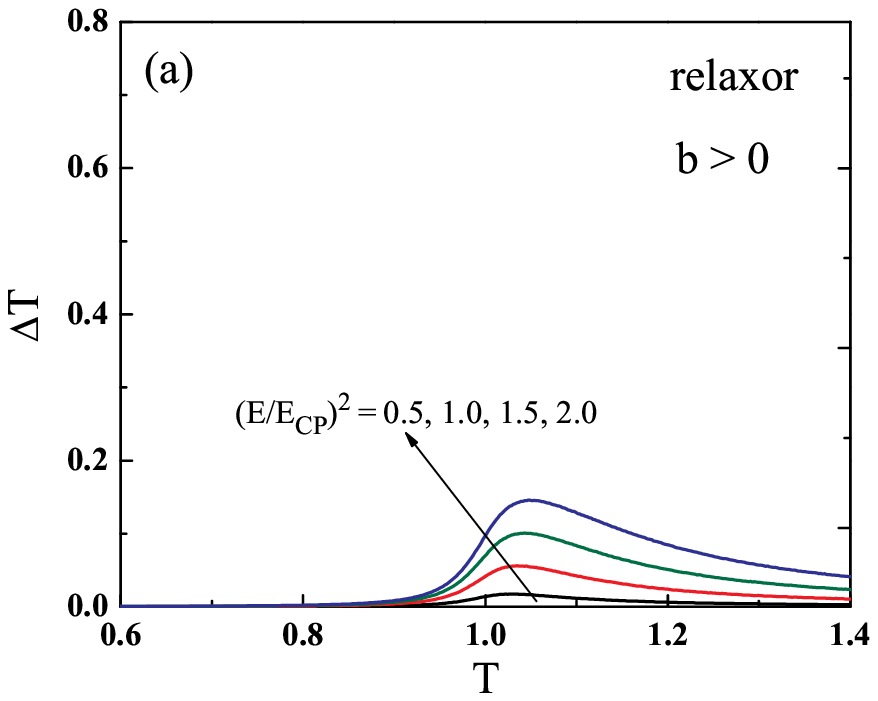}} &
\resizebox{20.4pc}{!}{\includegraphics{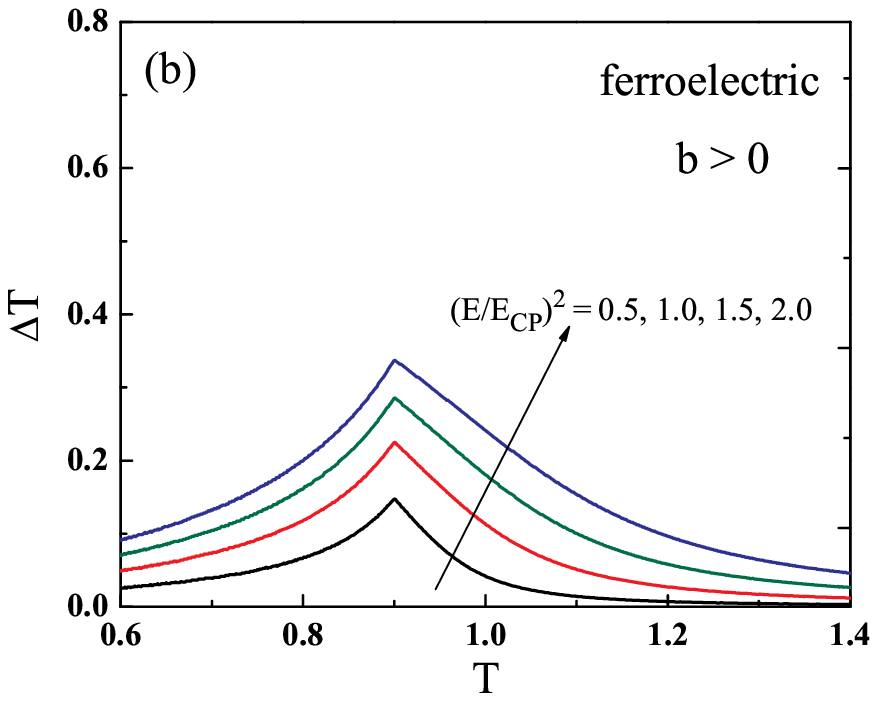}}\\ 
\end{tabular}
\caption{(a) Calculated temperature dependence of $\Delta T$ in 
a relaxor with $b>0$ and four values of electric field $E$, 
as indicated.
(b) Same, but for a ferroelectric.}
\label{pirc_fig2}
\end{figure*} 
\end{widetext}


\begin{widetext}
\begin{figure*}
\begin{tabular}{cc} 
{\large (a)}&{\large (b)}\\
\resizebox{20.4pc}{!}{\includegraphics{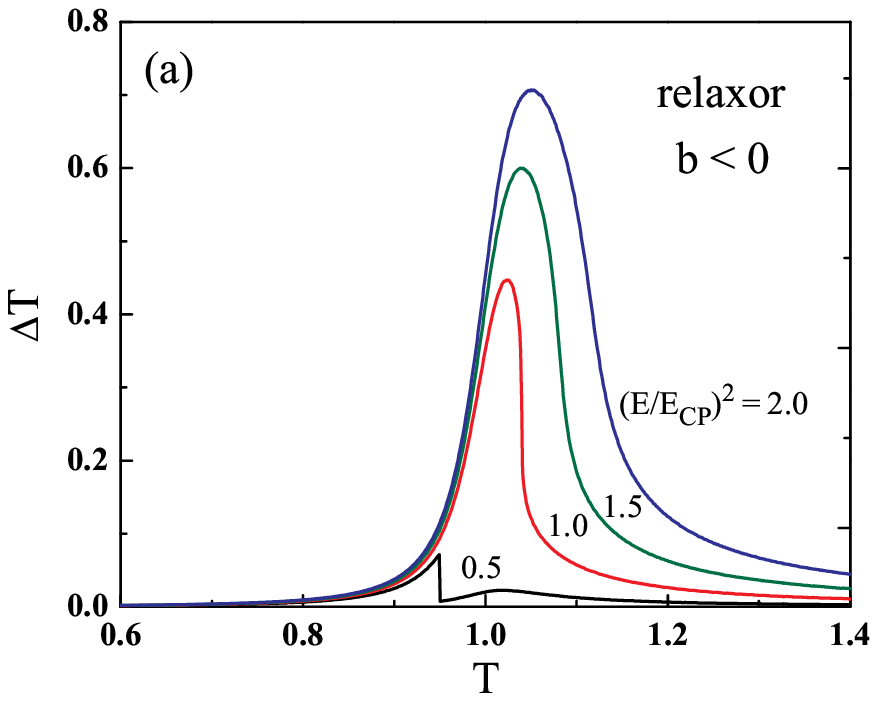}} &
\resizebox{20.4pc}{!}{\includegraphics{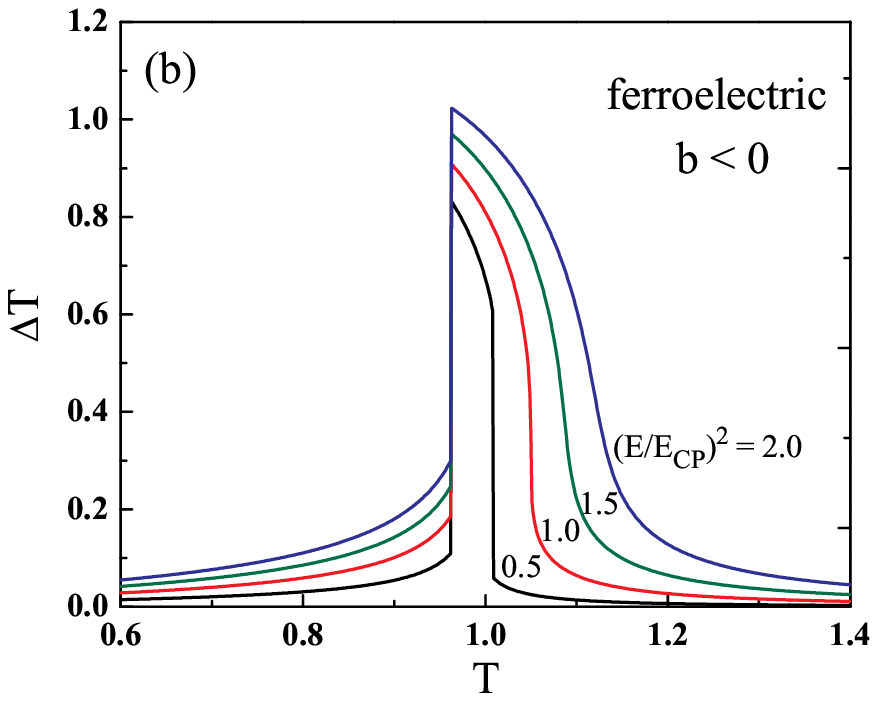}}\\ 
\end{tabular}
\caption{(a) Calculated temperature dependence of $\Delta T$ 
in a relaxor with $b<0$ and four values of electric field $E$, 
plotted on the same vertical scale as in Fig.~1. The critical 
point lies at $T_{CP}=1.0398$ and $E_{CP}=0.0438$.
(b) Same, but for a ferroelectric with $T_{CP}=1.05$ and 
$E_{CP}=0.0438$, on different vertical scale. 
Note the first order phase transition at $T_1=0.965$.}
\label{pirc_fig3}
\end{figure*} 
\end{widetext}


\begin{widetext}
\begin{figure*}
\begin{tabular}{cc} 
{\large (a)}&{\large (b)}\\
\resizebox{20.4pc}{!}{\includegraphics{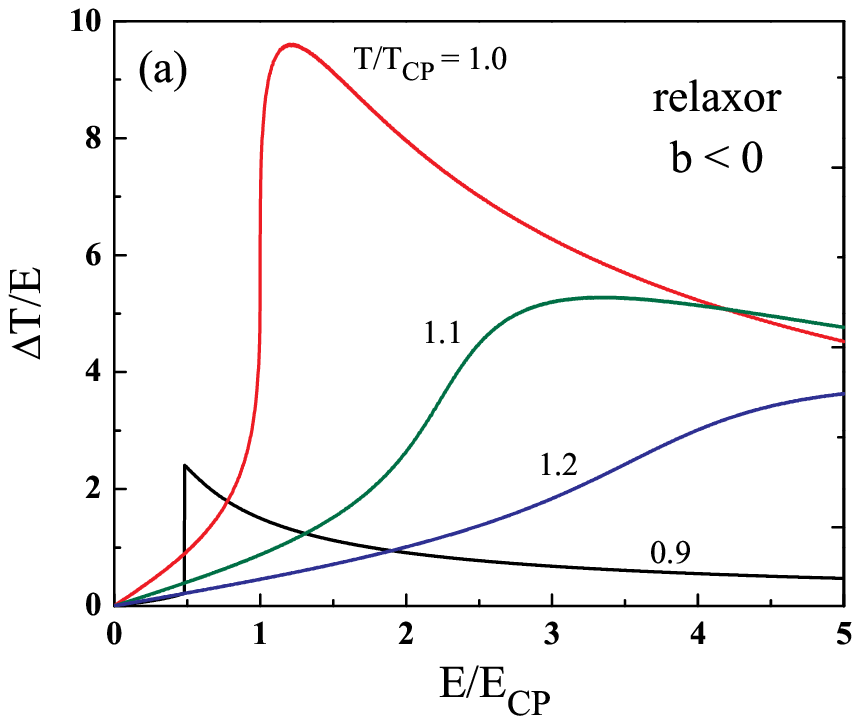}} &
\resizebox{20.4pc}{!}{\includegraphics{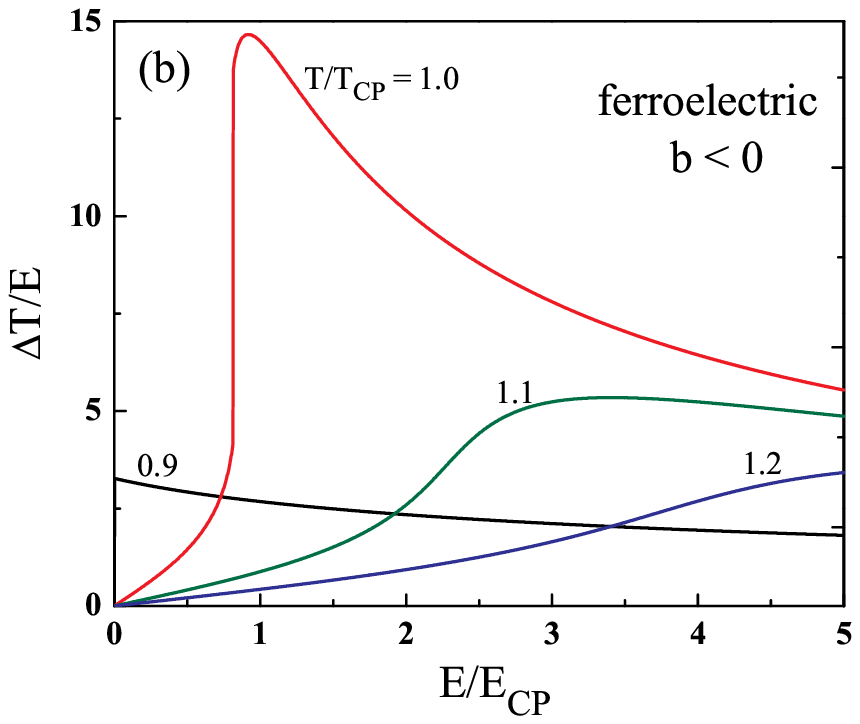}}\\ 
\end{tabular}
\caption{(a) Calculated field dependence of ECE efficiency
$\Delta T/E$ in a relaxor with $b<0$, at four temperatures
close to critical temperature $T_{CP}$. 
(b) Same, but for a ferroelectric, with different vertical scale.}
\label{pirc_fig4}
\end{figure*} 
\end{widetext}


\end{document}